\documentclass{svproc}
\usepackage{graphicx}%
\usepackage{multirow}%
\usepackage{amsmath,amssymb,amsfonts}%
\usepackage{mathrsfs}%
\usepackage[title]{appendix}%
\usepackage{xcolor}%
\usepackage{textcomp}%
\usepackage{manyfoot}%
\usepackage{booktabs}%
\usepackage{algorithm}%
\usepackage{algorithmicx}%
\usepackage{algpseudocode}%
\usepackage{listings}%
\usepackage{subcaption}
\usepackage{svg}
\usepackage{microtype}
\usepackage{multicol}
\usepackage[bottom]{footmisc}
\usepackage{url}

\begin{document}
\mainmatter              
%
\title{An Information-Theoretic Model for the Evolution of a Social Network}

\titlerunning{Entropy Driven Network Evolution}  
%
\author{Johnathan Brink \and Lewis Mitchell \and 
Mengbin Ye}
\authorrunning{J. Brink et al.}
\institute{Adelaide Data Science Centre, School of Mathematical Sciences, Adelaide University, Australia\\
\email{johnathan.brink@adelaide.edu.au}
}

\maketitle              

\begin{abstract}
Echo chambers and filter bubbles provide a framework to view communication and connection in a social network age. Such systems are characterised by isolated communities regurgitating information which reinforces prior beliefs. Theoretical analysis of echo chambers is suited for a network science approach to understand the mechanics of information influences between users. Here, we propose an evolutionary network model that extends the Quoter Model, rewiring edges according to cross-entropy between text-generating nodes. Parameters characterise binary reinforcement vs exploration node intents, determining whether a node will seek high or low novelty neighbours, measured by cross-entropy. Through an empirical simulation study, we observe emergent model attributes. Specifically, we compare to a Watts--Strogatz model, highlighting parameter regions of high clustering and hyper-exploration. Further, we identify a quality of disjoint communities akin to rudimentary echo chamber structures. These findings underscore user behaviours that drive echo chamber entrenchment, contextualising future social network analysis.
\keywords{Network Evolution Methods, Cross-Entropy, Quoter Model, Selective Exposure}
\end{abstract}

\section{Introduction}
{
Echo chambers and filter bubbles represent a defining challenge of the modern social media landscape, fundamentally altering social and political relationships by isolating communities within self-affirming information loops \cite{echoch,vic}. Understanding how influence relationships develop and converge within these social structures is crucial, as the continuous reinforcement of pre-existing beliefs drives partisan polarisation and algorithmic entrenchment \cite{macy}.

The formation of these systems has been explored and discussed extensively through empirical observation, highlighting that users frequently engage in selective exposure to like-minded individuals, while algorithms further direct them toward ideologically aligned information \cite{selexp,filtbub}. Previous studies have considered network structure and user content consumption patterns on these platforms, with a focus on identifying their role in political discussion \cite{repub,benecho}. Further, there have been efforts of opinion dynamic evolutionary models to capture echo-chamber behaviour \cite{kan}. 

Despite widespread documentation of online echo chambers, a gap exists regarding analytical and agent-based simulation approaches that can be used to model dynamic information flow. Little research mathematically bridged user seeking mechanics (whether it be seeking specific users or content) with the ongoing, macroscopic evolution of the underlying network structure itself. Addressing this gap is necessary to map out exactly how localized human interactions and varying degrees of reinforcement preference shape the formation of group-isolated networks. Our foundation for modelling localized interactions is the ``Quoter Model'', which captures information flow and influence through text generation. The strength of this flow is dictated by quoting probability $q$ \cite{quoter,pond}. This model allows inter-node influence through information diffusion as pieces of text propagate between neighbours via quoting mechanics. Through these mechanics, node selection bias is formed solely by content exposure, representing the influence that occurs through social interaction.

In this paper, we propose an information-theoretic model that extends the Quoter framework to incorporate a time-varying network structure shaped by cross-entropies between nodes \cite{Konto}. Specifically, we introduce a novel evolutionary mechanism where network edges are dynamically rewired according to the cross-entropy between text-generating nodes. Through parameter $r\in[0,1]$, we designate user behaviour as either reinforcement-seeking (low novelty) or exploration-seeking (high novelty), simulating the network's structural progression under varying behavioural assumptions, including the ``trench warfare'' dynamics of online discourse \cite{trench}. Our core contribution demonstrates that there exist combinations of quoting probability and reinforcement behaviour that lead to distinct parameter regions characterized by high clustering or hyper-exploration. Through rigorous comparison against a Watts--Strogatz model, our numerical simulations reveal the organic emergence of disjoint, highly clustered communities that structurally mirror rudimentary echo chambers \cite{WS}.

Our methodology follows from previous efforts to define and study adaptive networks, in which the network structure evolves by some adaptive mechanism. Specifically, our model can be considered a co-evolutionary network wherein both the network topology and individual node states evolve interdependently \cite{gross2008}. Classical coevolutionary models consider homophily as a driving mechanism, often measured as normalised or discrete opinion states which dictate rewiring \cite{baumann2020}. As an example, a highly influential co-evolutionary model by Holme \& Newman \cite{holme2006} captures network adaption under an opinion dynamics framework, utilising control parameters to capture decisions between rewiring and change of opinion. Analysis of this model found a parameter dependent threshold determining whether the network reaches a global consensus or fragments into isolated communities, a result innate to many adaptive networks \cite{kozma2008} and similar to our information-theoretic model.

The remainder of this paper is structured as follows. Section \ref{background} formally presents the Quoter Model and outlines our entropy-driven rewiring. Section \ref{results} details the empirical simulation results, comparing emergent topological metrics against standard network generating algorithms. Finally, Section \ref{conclusion} offers concluding remarks and directions for future theoretical extensions.}

\section{Model}\label{background}
In this section, we present the Quoter model with evolutionary entropy-driven rewiring. First, we introduce the notion of text entropy and the fundamental Quoter model. Then, we present the entropy-driven rewiring mechanism.

\medskip

\subsection{Quoter Model}\label{subsec2}
The original Quoter model was proposed to capture information flow and influence via text-based conversations between users in an online social network~\cite{quoter,pond}. Content is created by users through one of two mechanisms: i) original thought, or ii) quoted ideas from another user. 

\begin{figure}[h]%
\centering
\includegraphics[width=0.8\textwidth]{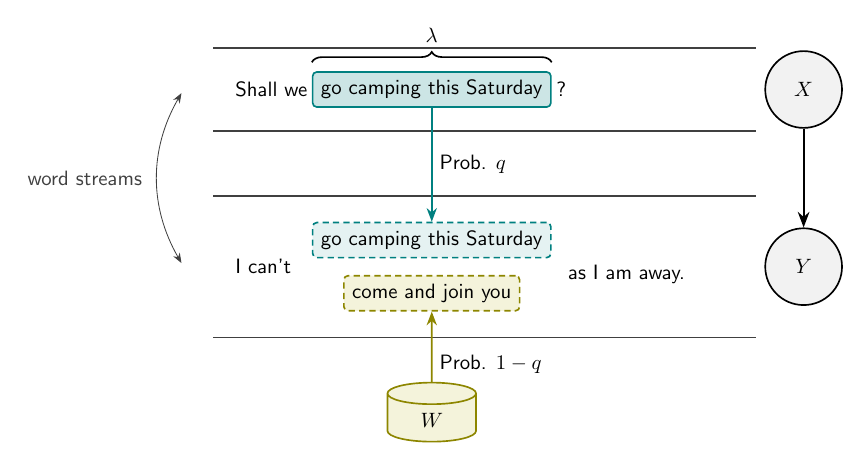}
\caption{An example of the text generation from $Y$, given neighbour $X$. A string of length $\lambda$ is quoted from $X$ with probability $q$ or from a dictionary distribution, $W$, with probability $1-q$}\label{quoter example}
\end{figure}

The model is defined on an undirected graph $G = (V,E)$, with $N\geq2$ nodes captured by the node set $V$ and edge set $E \subseteq V\times V$. 
Nodes are indexed from $1, \hdots, N$, representing people or users in the network. Each node has associated with it a word stream, which is the collection of text that user has generated in the past. This represents content such as social media posts, tweets, and comments. An edge $(i,j)$ indicates that nodes~$i$ and $j$ have access to each other's word stream, and $j$ is said to be a neighbour of $i\in V$ (and vice versa). 
At each discrete timestep $t\in \mathbb N$, a single node $i$ is chosen uniformly at random, and will then generate a length, $\lambda\sim L$, amount of words, where $L$ is an integer distribution to be specified. Node $i$ generates words under two possible mechanisms based on a quoting probability $q\in[0,1]$ (see Fig. \ref{quoter example}):  
\begin{enumerate}
    \item With probability $1-q$, $\lambda$ words are drawn from a word distribution, $W$, defined below.
    \item Alternatively, with probability $q$, a neighbour of $i$, say $j$, is picked uniformly at random and $\lambda$ words are quoted in sequence, starting from a random index in $j$'s word stream. For a word stream of length $m$, this index is chosen in the range $[ 1, m-\lambda]$ to preserve the distributed length, ensuring a quoted segment of $\lambda$ words.
\end{enumerate}


There are various decisions to be made around this model, including our distributions, $L$ and $W$. As is consistent with similar literature, a reasonable choice for $L$ is the Poisson distribution, preserving the integer property of $\lambda$ \cite{quoter}. For $W$, it is common to use the Zipf distribution \cite{zipf}.

\subsection{Entropy Estimates}
Crucial to the framework of our evolutionary network model is understanding influence and information between nodes. In our framework, this takes the form of text entropy. The entropy, $h$, of a text is denoted as the number of additional units of information needed on average to determine the next word given all past words \cite{Entropy}. This can be generalised to a measure of cross-entropy between the text streams of two nodes, $A$ and $B$, given by

\begin{equation}
    h_\times(A|B) = \frac{T_A \log_2 T_B}{\sum_{i=1}^{T_A} \Lambda_i(A|B)}\,,
    \label{crossent}
\end{equation}
where $T_A$ and $T_B$ are the respective text lengths of $A$ and $B$ \cite{Konto}. Here, the match length $\Lambda_i(A|B)$ is the shortest substring in $A$ starting at index $i$ (generated at time step $t$) that has not previously appeared in $B$ before $t$. Intuitively, this metric measures $B$'s influence on $A$ through the average amount of additional information needed to encode the next word of $A$ given all information in $B$.

Previous research has focused on how information is diffused throughout a network under a Quoter model framework \cite{quoter,cross,pond}, the main result being that cross-entropy grows in direct relation to a node pair's path length and its degree.

\subsection{Cross-Entropy Driven Rewiring}

We now propose an extension to the Quoter model by introducing an evolutionary mechanism for rewiring edges, driven by a user's desire to maximise or minimise cross-entropy. In the online social media context, this captures users unfollowing and following each other's accounts. The framework for this evolution is described  below.
\begin{enumerate}
    \item Pick a node uniformly at random from the network, say node $i$.
    \item Compute cross-entropy between $i$ and every other node (complexity $\mathcal{O}(N)$).
    \item Select an edge containing node $i$, say $(i,j)$ and \textit{rewire} by severing this edge and connecting $i$ to a different node other than $j$.
\end{enumerate}
Importantly, rewires will ensure a conservation of edges, such that the total number of edges in the network remains constant over time. This provides a consistent and fair comparison with which to view results, as network metrics cannot be influenced by the increase or decrease of edges.

The frequency of these rewires relative to the text generation process must also be considered. We define new parameter $\theta\in\mathbb{N}$ such that a rewire occurs every $N \theta$ timesteps. Due to the nature of the Quoter model, this ensures that every node has on average $\theta$ text generation occurrences between rewires. As entropies will converge over time, an increase in $\theta$ implies a stronger influence to be borne between new node pairs as rewires occur.

Network dependence on time implies that quoting mechanics are not consistent between nodes in that a node's neighbours change over time. As rewires occur, quoting is promoted between a new node pair and ceases from a previous pair.

We explore a range of potential drivers for a user to rewire (which in context means changing which other users they want to observe word streams from).
We posit that users will either rewire based on a \emph{reinforcement} or \emph{exploration} trading mechanism. Here, reinforcement trading is a tendency to rewire to nodes with a lower cross-entropy and hence a more similar past word stream. Conversely, an exploration trade rewires nodes to more dissimilar non-neighbours with higher cross-entropies. We model this via the parameter, $r\in[0,1]$, such that $r$ and $1-r$ are the probabilities that a chosen user conducts a reinforcement trade and exploration trade, respectively.

We also introduce a weighted distribution to the choice of which node to rewire towards, using a noisy (or bounded) rationality interpretation to the process, inspired by game-theoretic mechanisms of decision-making \cite{boundrat1}. Specifically, we utilise the softmax function \cite{softmax} which normalises a vector of choice scores, $\vec{z} \in \mathbb R^K$ such that each choice, $z_i,$ $ i\in\{1,\dots,K\}$, is assigned a normalised weight:
\begin{equation}
    \sigma(\vec{z},\beta)_i=\frac{\mathrm{e}^{\beta z_i}}{\sum_{j=1}^K\mathrm{e}^{\beta z_j}}\,.
\end{equation}
Here, $\beta \in [0,\infty)$ is a measure of rationality, whereby increasing $\beta$ corresponds to increasing rationality and towards picking the highest utility choice. The process becomes deterministic when $\beta \to \infty$ and uniformly random as $\beta \to 0$. 

Define $h^{i\sim j}_\times$ and $h^{i\nsim j}_\times$ as the cross-entropy computed between $i$ and a neighbour or non-neighbour $j$, respectively. Then, the vector probabilities of a rewire are governed by:
\begin{equation}\label{remove}
    \mathbb{P}(\text{disconnect } (i, j)) =r\sigma(h^{i\sim j}_\times,\beta)+(1-r)\sigma(-h^{i\sim j}_\times,\beta)\,,
    \end{equation}
    \begin{equation}\label{add}
    \mathbb{P}(\text{connect } (i, j)) =r\sigma(-h^{i\nsim j}_\times,\beta)+(1-r)\sigma(h^{i\nsim j}_\times,\beta)\,.
\end{equation}
We summarize this rewiring process in Algorithm \ref{rewire}. Roughly speaking, a rewire under this framework first computes the cross-entropies from $i$ to its neighbours and non-neighbours, forming the vectors $\vec{h}^{i\sim j}_\times$ and $\vec{h}^{i\nsim j}_\times$. Node~$i$ then decides whether to perform a reinforcement trade with probability $r$ or exploration trade with probability $1-r$. If a reinforcement trade is to occur, node~$i$ disconnects from a neighbour $j$ (with preference for higher entropy neighbours to be disconnected) while a non-neighbour $k$ is chosen to rewire to (with preference for connecting to non-neighbours with low entropies). In the case of an exploration trade, these preferences are swapped, with node $i$ seeking higher entropy connections and disconnects from lower entropy neighbours.

\begin{algorithm}
\caption{Rewire}\label{rewire}
\begin{algorithmic}[1]
\Require Choice of $r\in[0,1]$ and $\beta\in[0,\infty)$
\Ensure $i$ has degree, $d$, s.t. $0<d<N-1$.
\State Compute cross-entropy between $i$ and its neighbours, $\vec h^{i\sim j}_\times$, and non-neighbours, $\vec h^{i\nsim j}_\times$ 
\State $r_{\mathrm{sample}} \sim U(0,1)$ 
\If{$r_{\mathrm{sample}} < r$}
    \State $n \sim \sigma(\vec h_\times^{i\sim j},\beta)$
    \State $s \sim \sigma(-\vec h_\times^{i\nsim j},\beta)$
\ElsIf{$r_{\mathrm{sample}} \geq r$}
    \State $n \sim \sigma(-\vec h_\times^{i\sim j},\beta)$
    \State $s \sim \sigma(\vec h_\times^{i\nsim j},\beta)$
\EndIf
\State Replace edge $(i,n)$ with edge $(i,s)$
\end{algorithmic}
\end{algorithm}

{It is important to consider the complementary contexts behind the two rewiring mechanisms. Here, a reinforcement trade (from high to low entropy) denotes a motivation grounded in social consistency. This is a key behaviour behind content-seeking in which users seek information which coincides with prior beliefs, sociologically thought to be a driver of ``echo chambers'' \cite{selexp}. In contrast, an exploration trade (from low to high entropy) acts as a novelty-seeking mechanism. A rewire under this structure causes nodes to seek a dissimilar non-neighbour, a concept established in observed social-media behaviour in which users may search for converse views for a variety of reasons. This includes natural curiosity as well as a hunt for inflammatory content to engage in online ``trench warfare'' \cite{trench,nov}.

The primary novelty of this paper is the introduction of the aforementioned evolutionary network dynamics driven directly by information-theoretic influence. Unlike static models of information diffusion, nodes in this system adaptively rewire their connections based on cross-entropy, simulating users actively seeking out either novel or similar information streams. This ongoing interplay of exploration and reinforcement trades creates a network topology that organically evolves over time. Furthermore, our exploration parameter, $r$, permits contention between possible user motivations, allowing an understanding of how resultant networks may form from a reinforcement vs. exploration approach.
Hence, tracking the structural progression of the network is essential to understanding how initial conditions and quoting behaviours generate emergent phenomena.}

\subsection{Simulation Framework}\label{sim}
{
Due to the complex nonlinearities of the model, we use Monte Carlo simulations to characterise its emergent phenomena. 
In keeping with the properties of language, we assumed a Zipfian word distribution with a vocabulary size $z=1000$ and $\alpha=1.5$. Further, we assumed the text length distribution to be Poisson with distribution parameter, $\lambda=3$. 

At each rewire step, we need to determine $\mathcal{O}(N)$ cross-entropies. Referring to Eq.~\eqref{crossent}, we see that the match length complexity, and thus computation, grows at $\mathcal{O}(t^2)$ as text strings increase. Fortunately, previous effort has been made for efficient C++ match-length determination \cite{LCS}. This reduces compute time significantly, but simulations on large networks and many timesteps are slow.

Simulations explored parameter ranges $q,r\in\{0,0.05,0.1,\hdots,1\}$ and $\beta\in\{2^0,2^1,\hdots,2^7\}.$ These choices of $\beta$ allow for an exploration across the rationality spectrum with $\beta=2^7$ corresponding to a closely deterministic system. Furthermore, we set $N=100$ nodes, chosen as  prior exploration revealed that a network of this size captured features of larger scale networks while still being low enough to satisfy $\mathcal{O}(N)$ computational feasibility. Following this, a time horizon of $T=3\times 10^6$ was set as network models of this size were found to have converged in preliminary simulations. We set $\theta=100$ for an inter-rewire time of $N\theta=10,000$ and hence $T / (N\theta)=300$ rewires occur throughout the simulation. Under this condition, each node generated text on average $100$ times between rewires, a choice that allowed substantial opportunities of quoting between node pairs, converging their respective cross-entropies \cite{quoter}. Each parameter set was simulated across 10 independent trials, allowing for averaging of metrics.

We considered Watts--Strogatz (WS), Erdos--Renyi (ER), and Barabasi--Albert (BA) for network initialisation. As our model follows a rewiring structure, rather than node or edge generation, we decided to start from a ring lattice equivalent to a WS graph with rewiring probability $p=0$. This allows for direct comparison between our evolving network obtained through cross-entropy-driven rewiring, with the WS model that has random rewiring when $p > 0$, approaching an ER network as $p\to1$. For all trials in our experimentation, we initialise the ring lattice with degree, $K=4$.

As well as the final network (i.e. at the end of the simulation window), a variety of metrics were collected throughout the simulation to better describe the model's behaviour. Particularly: global clustering coefficients, $C$, average path lengths, and Louvain modularities, capturing the network structures  \cite{modularity}.

\section{Results}\label{results}

We now explore how network metrics for our model depend on parameters $(q,r,\beta)$ from Sect. \ref{sim}, and compare to a standard WS network~\cite{WS}.

\begin{figure*}[h]%
\centering
\includegraphics[width=0.8\textwidth]{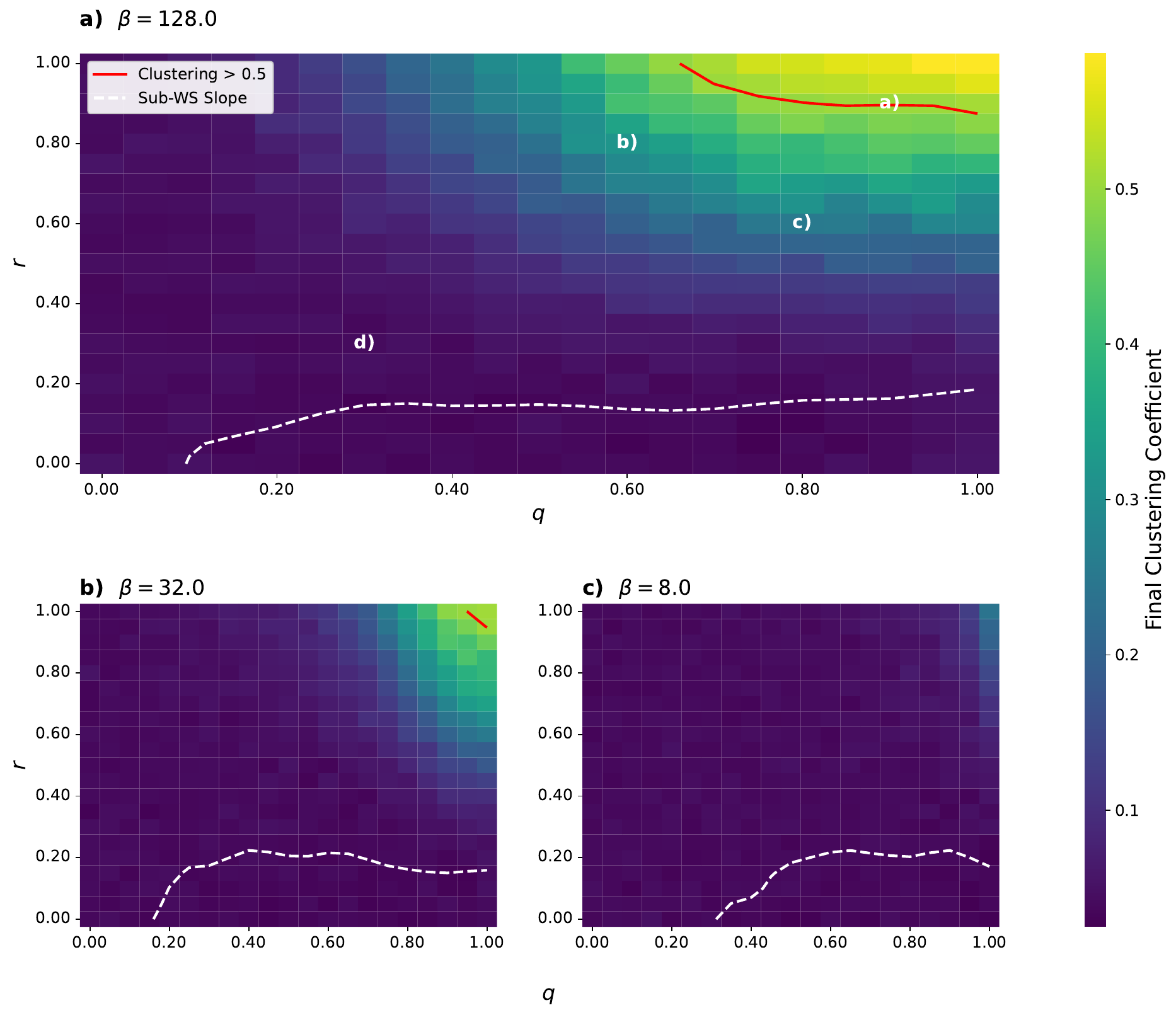}
\caption{Global clustering coefficient heatmaps at simulation termination across three values of $\beta$ with axis parameters of $q,r$ (see Sect. \ref{sim}). Level curves indicate the regions of $C>0.5$ (red) and steeper than WS slopes (white). Letter markings on a) indicate the locations of network examples in Fig. \ref{networks} \vspace{-8pt}}\label{heatmap}
\end{figure*}

Fig. \ref{heatmap} displays $(q,r)$ heatmaps of final simulation global clustering coefficients, $C$, across a range of $\beta$ values. Each cell is averaged across 10 trials, showing a clear diffusive pattern with positive correlations across both $q$ and $r$. Highly clustered communities emerge as $q$ and $r$ increase, where quoting behaviour is frequent and nodes are attracted to similar non-neighbours. This pattern weakens at lower $\beta$ values, leading to smaller $C$. We therefore only consider the near-deterministic $\beta=128$ case for the remaining discussion to better explore the pattern between $C$ and $(q,r)$.

Note that the $C$ for the initial ring lattice network is 0.5 and under some conditions the rewiring increases $C$ from this baseline, an occurrence in stark contrast with the WS model, where the clustering coefficient always decreases as edges are rewired to form shortcuts. This region, marked in red within Fig.~\ref{heatmap}, exists at large $q,r,\beta$ values, requiring strong influence from quoting ($q$) and high reinforcement ($r$) with high rationality ($\beta$).

\begin{figure*}[h]%
\centering
\includegraphics[width=\textwidth]{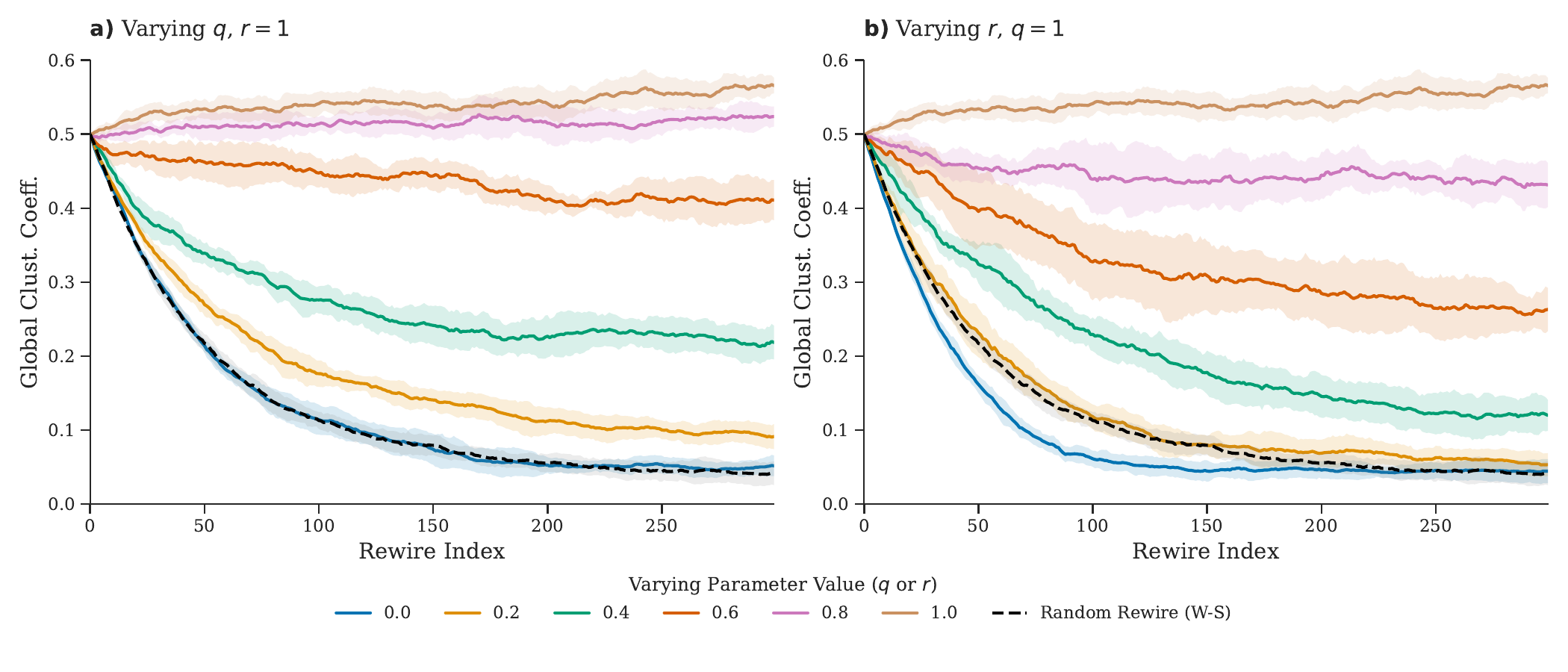}
\caption{Comparison of global clustering coefficients over time (rewires) when varying parameters $q$ (a) and $r$ (b). All results are averaged across 10 trials to include 95\% confidence intervals. When $q,r$ not varied, parameter is set to 1. All simulations have parameters detailed in \ref{sim} and $\beta=128$ \vspace{-8pt} }\label{evolve}
\end{figure*}

However, $q$ and $r$ impact evolution of the network unequally. Fig. \ref{evolve} allows for better understanding of how the network may evolve over time as users repeatedly rewire in comparison to a WS global clustering coefficient baseline of random rewires (black dashed line). We see similar trends between $q$ and $r$ with some notable yet subtle differences. As $q\to0$ with $r=1$, the model approaches the WS baseline, an intuitive result due to information flow via quoting halting at $q=0$ and rewires essentially becoming random. However, as $r\to0$ with $q=1$, the decrease in $C$ is at a greater rate than the baseline, indicative of early rewires cutting across longer path lengths. We call this behaviour where a sub-WS slope occurs ``hyper-explorative'' as it indicates drastic but deliberate rewires across the lattice, decreasing $C$ by almost always introducing a new open triangle. To better identify this behaviour, we highlight this hyper-explorative region in white within Fig. \ref{heatmap}. Interestingly, this occurs at the low $r$ range but also requires sufficiently large $q$. Intuitively, this is because nodes require influence (quoting) to identify novelty, a feature of the interplay between the parameters. As $q$ decreases, causing a global increase in cross-entropy due to less influence between the nodes, the selection when exploring becomes less specific. Again, at $(0,0)$, the model essentially returns to random rewiring.

We also observe interesting patterns around the varied parameters in Fig. \ref{evolve}. Clearly, there exists some threshold of separation between networks with $C$ that quickly reaches a high limiting value, and those which will converge to a structure similar to an Erdos--Renyi network with low $C$ as $T\to\infty$. For example, in Fig. \ref{evolve} a), we see the evolution for $q=0.8$ stabilise just above $C=0.5$ while at $q=0.6$, $C$ continues to decrease even after 300 rewires. In other words, there exists nuance around the termination choice of this model as extended $T$ values of the simulation in lower $(q,r,\beta)$ ranges will significantly decrease the clustering in the final networks.

\begin{figure*}[htbp]
    \centering
    \begin{subfigure}[b]{0.23\textwidth}
        \centering
        \includegraphics[width=\textwidth]{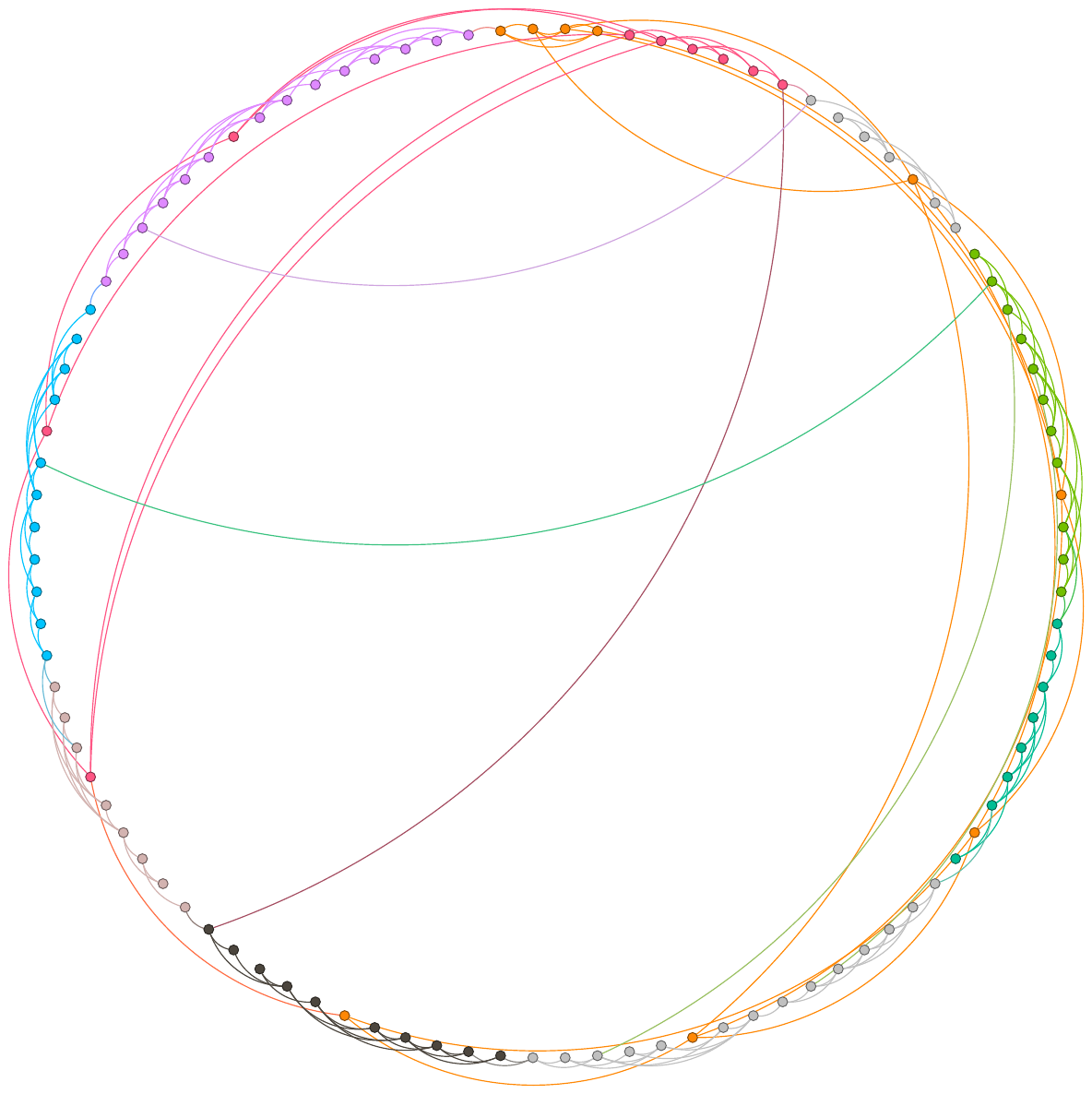}
        \caption{
        (0.9,0.9)
        }
        \label{fig:3_3}
    \end{subfigure}
    \hfill
    \begin{subfigure}[b]{0.23\textwidth}
        \centering
        \includegraphics[width=\textwidth]{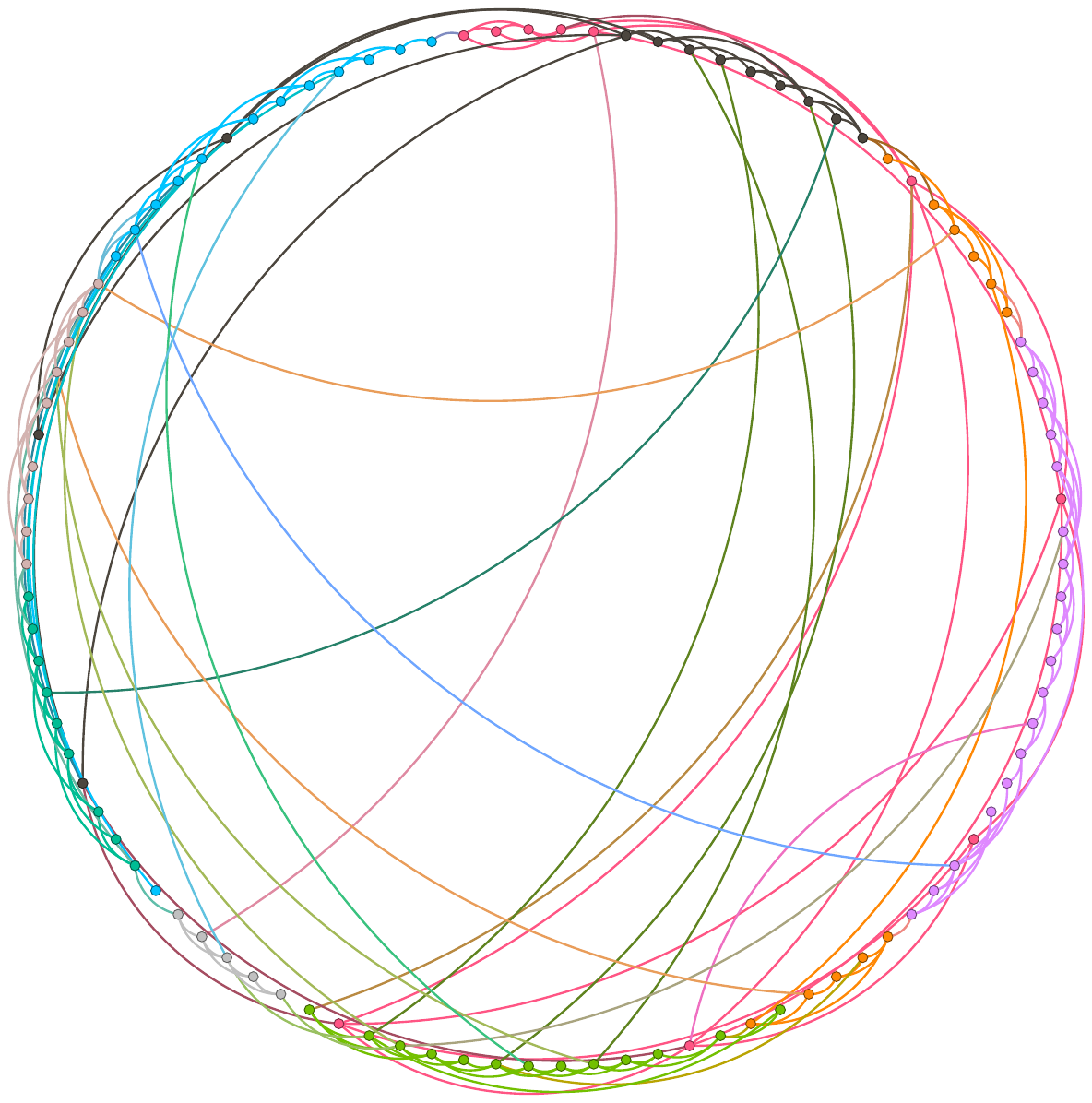}
        \caption{(0.6,0.8)}
        \label{fig:6_8}
    \end{subfigure}
    \hfill
    \begin{subfigure}[b]{0.23\textwidth}
        \centering
        \includegraphics[width=\textwidth]{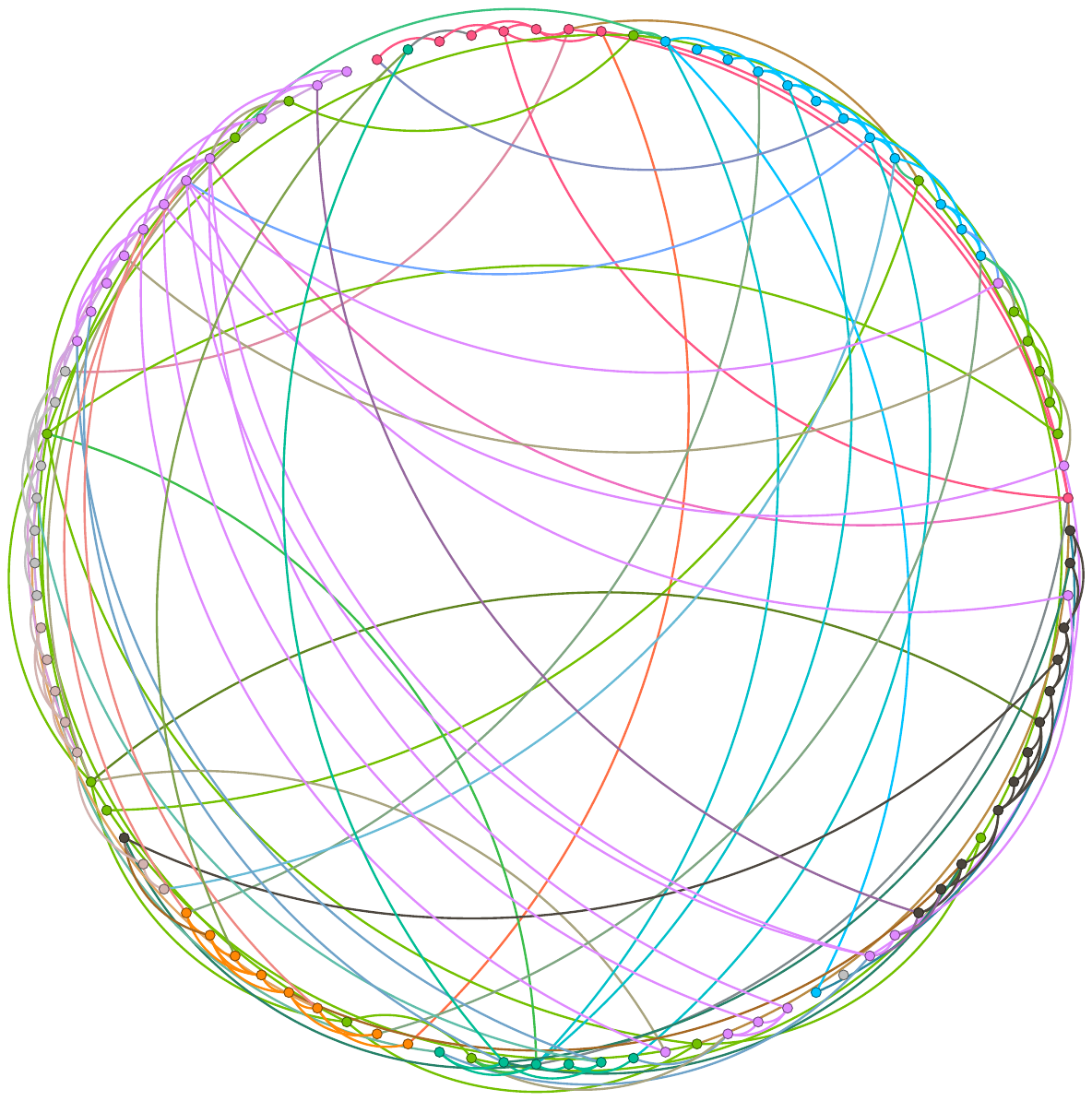}
        \caption{(0.8,0.6)}
        \label{fig:8_6}
    \end{subfigure}
    \hfill
    \begin{subfigure}[b]{0.23\textwidth}
        \centering
        \includegraphics[width=\textwidth]{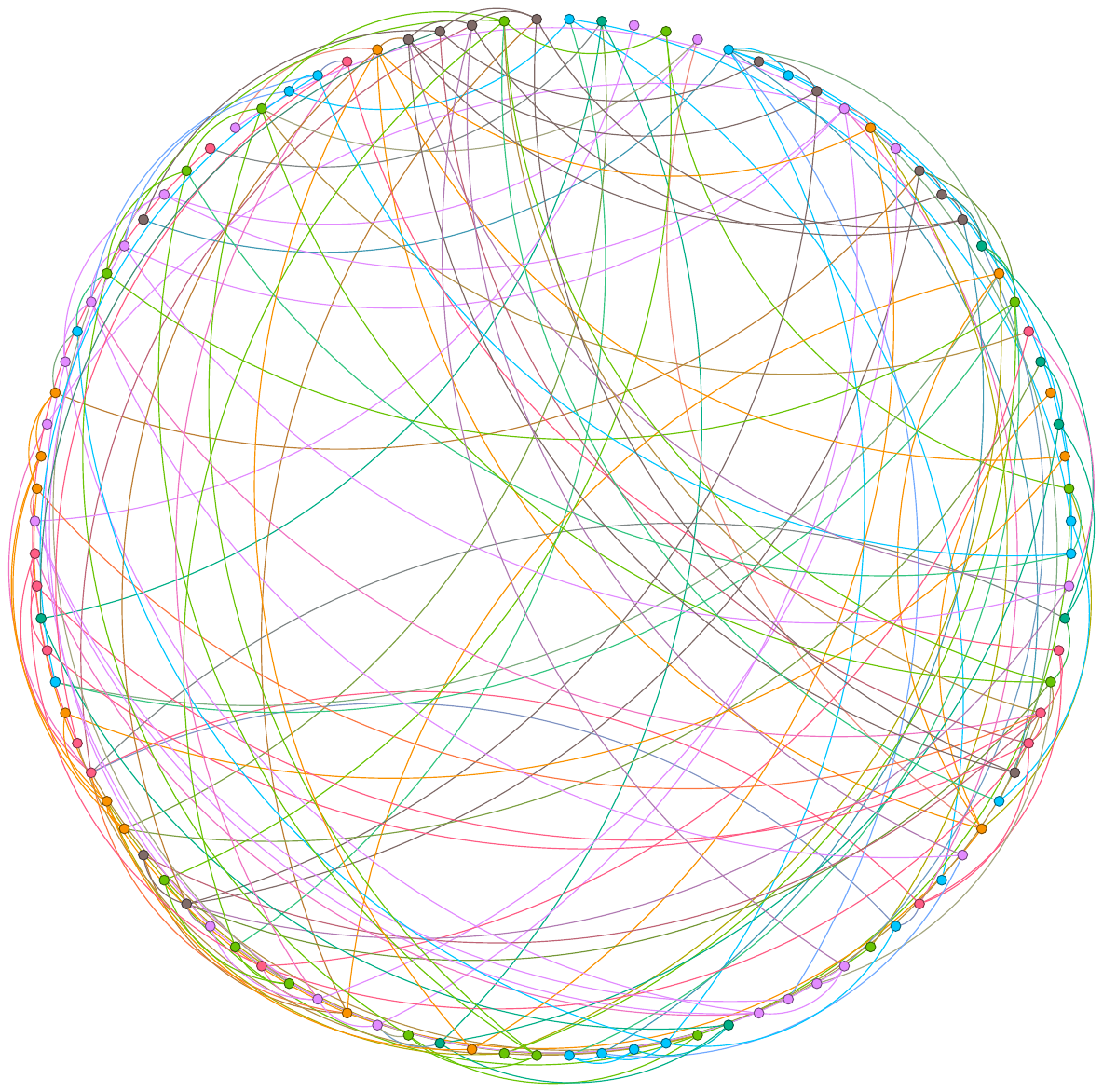}
        \caption{(0.3,0.3)}
        \label{fig:9_9}
    \end{subfigure}
    \caption{Network examples in ($q,r$) space at termination of simulation. Nodes and edges coloured by modularity classes, calculated by Louvain's method with resolution of 1. All simulations have parameters detailed in \ref{sim} and $\beta=128$ \vspace{-8pt}}
    \label{networks}
\end{figure*}

Fig. \ref{networks} shows some examples of networks after 300 total rewires. We choose four points in $(q,r)$ space for this comparison, marked above in the Fig. \ref{heatmap} heatmap. We measure and partition by the Louvain modularity of these networks, highlighting the distinct groups within. These examples qualitatively demonstrate the main features of our model. 

Firstly, consider the example at the highest values, $(0.9,0.9)$. Compared to the other networks, there are notably fewer lattice shortcuts, indicated by the sparseness in the middle of the circular network. There also exists strong grouped modularity classes around this lattice. Now, consider that this network has $C>0.5$ despite several existing shortcuts. This is due to a high clustering strength within the observable classes. That structure appears to divide into distinct communities, with a high density of edges existing within groups of adjacent nodes on the lattice. Meanwhile, connections between these communities are sparse, with information being diffused by the quoting mechanism much less efficiently. These communities persist but weaken within Fig. \ref{networks} b) and c). However, d) shows the collapse of this high clustering structure with a strong decrease in adjacent lattice neighbours.

Given the context of the Quoter Model, the distinct communities emerging here are representative of echo chambers, characterised by strong ties to similar values (word streams) quoting between users and an isolation from those dissimilar in utterances. Due to information diffusion by our quoting mechanism \cite{quoter}, similarity approximately maps to proximity within the circular lattice. Similarly to $C$, average path lengths increase with $q$ and $r$, furthering this isolation. Additionally, the modularity metrics of this network yield a similar pattern to Fig. \ref{heatmap} a), indicating more identifiable groupings at higher $(q,r)$ values. These results are omitted due to space limitations.

This model exhibits community formation with strength dictated by parameters $(q,r)$. While more visually obvious with the WS initialisation, this modularisation trait is not unique to initial network structure, with similar communities and high $C$ emerging when commencing from Barabasi--Albert or Erdos--Renyi networks. Furthermore, to contextualise our parameters, strong community groups require both high susceptibility to influence ($q$) and active reinforcement by seeking similarity ($r$). These trends are consistent with research on behaviour in online spaces. For example, those with higher political engagement are more likely to select content that aligns with pre-existing views \cite{repub}. Moreover, users in echo chambers typically exhibit high political activity \cite{Twitter}. While the existence and true definition of echo chambers remains contested, in real networks high $q$ can be associated with high $r$ and vice versa.

\section{Conclusion}\label{conclusion}
We have introduced a novel network evolutionary model based on inter-node influence, where quoting facilitates edge rewires. Using text cross-entropy to capture the influence between nodes, our model enables users to adapt network connections based on a desire to either maximize or minimize this entropy.

Exploring this model through Monte Carlo simulations yielded emergent, evolving network qualities, dependent on $q$ and $r$. Most interestingly, we identified that high quoting probabilities combined with high rates of reinforcement trades leads to highly clustered, distinct communities. These structures are tightly tied to similar word streams while remaining isolated from dissimilar communities, reflecting a rudimentary echo chamber. 

{
There exists a range of future work such as further exploring the boundaries within the $(q,r,\beta)$ regions of $C$ convergence referenced in Sect. \ref{results} as well as further examining the effects of initial parameters. We could also reassess some defining assumptions of our model, resulting in alternate node behaviour. One such change is limiting the memory of nodes' word streams via time horizons to better capture user behaviour and strengthen the influence between new node pairs, reducing computational complexity and offering a model where node pair relationships have less tendency to stabilise. We may also consider another choice for measuring homophily or introducing some level of heterogeneous behaviour.

Future models could assign explorative and reinforcing nodes, redefining $r$ as a proportion, representing conscious selective exposure and inspecting differences between these behaviours. Lastly, this model lends itself to LLM application, replacing quoting with training. Simulating this may provide insights into the conditions that support model collapse in LLMs \cite{modcol}.}

 \bibliographystyle{spmpsci} 
 \bibliography{ref1} 
\end{document}